\documentclass[prb,twocolumn,showpacs,citeautoscript]{revtex4}

\usepackage{graphicx}
\usepackage{amssymb}
\usepackage{amsmath}

\newcommand{\ud}{\,{\mathrm d}}

\newcommand{\uS}{{\cal S}}
\newcommand{\uU}{{\cal U}}

\begin{document}

\title{A model for flexural phonon dispersion in graphite and graphene.}

\author{Konstantin L. Metlov}
\affiliation{Donetsk Institute for Physics and Technology NAS, Donetsk, Ukraine 83114}
\email{metlov@fti.dn.ua}
\date{\today}
\begin{abstract}
  A simple model for flexural phonons in graphite (and graphene, corresponding
  to the limiting case of infinite distance between carbon planes) is proposed, in which the local dipolar moment is assumed to be proportional to the curvature of the carbon sheets. Explicit expressions for dispersion curves with full account for the long-range dipolar interaction forces are obtained and fitted to the experimental data using a single adjustable parameter of the model. The parameter is expected to depend on the ground state configuration of molecular $\pi$-orbitals, the same both for graphite and for graphene. At decreasing carbon sheet separation (high pressures) the phonon spectrum displays instability, corresponding to the graphite to diamond transition. Being explicitly based on the local dipolar moments, the proposed simple model may prove useful for considering electron-phonon interaction.
\end{abstract}
\pacs{81.05.U-, 63.20.-e, 63.20.dh}
\keywords{graphite, graphene, phonons, flexural phonons}
\maketitle

Recent interest to graphene, sparkled by its controlled production\cite{novoselov-Science2004} and promising electronic properties\cite{perebeinos_Nature2007}, produced demand for detailed study of all the related properties of layered carbon. While conduction of graphene is understood relatively well, there is still a need for a simple model for its mechanical properties\cite{perebeinos_PRB2009,KCC09}. In this paper such a model for out-of-plane (flexural) oscillations of carbon sheets is proposed with the emphasis on the long-range interactions between the induced electric dipoles. Due to orientation of the dipoles, the effect of such interactions on the flexural modes is the strongest.

The parameters of the model have clear physical meaning and the interaction of the induced dipoles fully accounts for their mutual orientation and distance. This is why the model can be expected to be transferable to consideration of out-of-plane motion of atoms in many (single-, many- and few-) layered carbon allotropes, differing only in the arrangement of atoms and, consequently, of the dipoles. As a test, the flexural phonon spectrum of graphite (and graphene, which is just a limiting case) is evaluated analytically and fitted to the experimental data from the literature. Despite its simplicity, the model correctly demonstrates an expected instability of graphite under pressure.

It is well known from Chemistry that carbon has valence of four and, in its layered form, makes three strong chemical covalent bonds to neighbouring atoms (carbon or others), called the $\sigma$-bonds. The remaining electron also participates in bonding, forming a weak $\pi$-bond, famous for making the electron delocalized across the whole molecule/crystal, which leads to many interesting properties of aromatic hydrocarbons, conduction of graphite and metallic-like conduction of graphene. The crucial importance of $\pi$-bonds in hydrocarbons was realized even before the H\"uskel model\cite{Huckel1}. These same bonds, as it will be seen from the next, define mechanical properties of layered carbon allotropes as well.

\begin{figure}[t]
  \begin{center}
    \includegraphics[width=8cm]{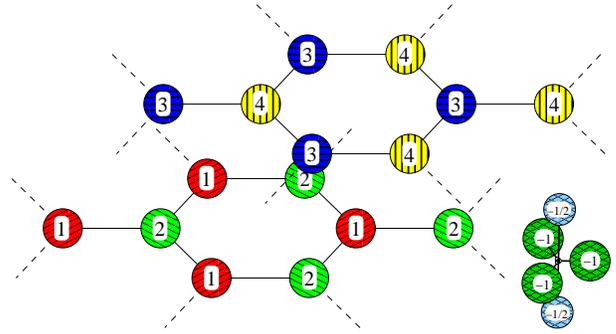}
  \end{center}
  \caption{Sublattices in the graphite lattice. Inset shows schematically the electron charge density around a single sp$^2$ hybridized carbon.}
  \label{fig:lattice}
\end{figure}

The electron density around a single sp$^2$-hybridised carbon atom is sketched as a ``ball and stick'' model on the inset in Fig.~\ref{fig:lattice}
. Balls represent centers of negative charge with $\sigma$-electrons shown as $-1$ charged balls in the horizontal plane, and the $\pi$-electron shown as two $-1/2$ balls above and below the plane. The thicker line between the $-1/2$ balls symbolizes electric connection between these charges, arising from the fact that they represent the same single electron. The whole picture is the result of momentum quantization (fixing the shape of electronic clouds) and simple electrostatic repulsion.

When three more carbon atoms are connected to the original one, additionally to forming $\sigma$ bonds, their $\pi$ clouds overlap, forming two ``seas'' of delocalized electrons above and below the plane, containing carbon nuclei. $\pi$  electrons spend half of their time above and half below the atom plane, and are free to move from one atom to the other. Evoking $\sigma$-$\pi$ separability and forgetting about $\sigma$ bonds we can imagine a single layer of carbon as three layers of charge: a layer of +1 (per atom) charges, representing the uncompensated charge of carbon ions, and two -1/2 charged layers of $\pi$ electrons on both sides of it.

Having this picture in mind, imagine that such a tri-layer carbon sheet is curved. Because both ``seas'' of electrons are connected (it is the very same electrons after all) the charge, pushed by the electrostatic repulsion, is free to redistribute from the contracting to the expanding side. This creates a local electric dipolar moment (interacting with similar dipolar moments across the layer), increases electrostatic energy of deformed electronic clouds and of the layer as a whole, producing the restoring force.

To describe this process mathematically, consider a graphite-like lattice, split into four independent hexagonal Bravais sublattices $1$-$4$, shown in Fig.~\ref{fig:lattice}. These sublattices are essentially the same, but only shifted with respect to each other, so that position of a lattice site, identified by three-dimensional integer vector $\vec{m}=[i,j,k] \in {Z}^3$ and number of the sublattice $l$, is
\begin{equation}
\label{eq:lattice}
\frac{\vec{r}^l_{\vec{m}}}{a} =
\underbrace{
\left[
\begin{array}{ccc}
\sqrt{3}/2 & -\sqrt{3}/2 & 0 \\
3/2 & 3/2 & 0 \\
0 & 0 & 2\alpha
\end{array}
\right]
\cdot
\left[
\begin{array}{c}
i \\
j \\
k
\end{array}
\right]
+
\vec{d}_l
}_{\vec{\rho}^l_{\vec{m}}} 
+
\left[
\begin{array}{c}
0 \\
0 \\
1
\end{array}
\right] u^l_{\vec{m}},
\end{equation}
where $a$ is the nearest neighbour distance in the lattice planes, $\alpha$ is dimensionless inter-plane distance (in units of $a$), $u^l_{\vec{m}}$ is out-of-plane displacement of atoms (in units of $a$), the matrix (denoted in the further text as $\mathbf{b}$) contains basis vectors of the lattice. The sublattice displacements (in units of $a$) are
\begin{equation}
\label{eq:shifts}
\vec{d}_{1} =
\left[
\begin{array}{c}
 0\\
0\\
0\\
\end{array}
\right], \,\,
\vec{d}_{2} =
\left[
\begin{array}{c}
0\\
1\\
0\\
\end{array}
\right],\,\,
\vec{d}_{3} =
\left[
\begin{array}{c}
 0\\
1\\
\alpha\\
\end{array}
\right],\,\,
\vec{d}_{4} =
\left[
\begin{array}{c}
0\\
2\\
\alpha\\
\end{array}
\right].
\end{equation}
All neighbours of an atom at sublattice $l$ belong to another sublattice $\tilde{l}$ (by definition: $\tilde{1}\equiv2$, $\tilde{2}\equiv1$, $\tilde{3}\equiv4$, $\tilde{4}\equiv3$). For the lattice (\ref{eq:lattice}) three nearest neighbours of an atom $\vec{m}$ on sublattice $l$ are the atoms $\vec{m}$, $\vec{m}^{+}_{l}=\vec{m}+\Delta_l[1,0,0]$, and $\vec{m}^{-}_{l}=\vec{m}+\Delta_l[0,1,0]$ of sublattice $\tilde{l}$, where $\Delta_l={\mathrm sign}(\tilde{l}-l)$; ${\mathrm sign}(x)$ is $1$ if $x\ge0$, $-1$ if $x<0$.

 There can be several definitions of local surface curvature, but, for the case of small deformations of the original lattice, all of them are essentially the same up to a constant multiplier. It is convenient to measure the curvature as a distance of the considered atom from the plane, defined by its three nearest neighbours. The normal to this plane at site  $\vec{m}$ of sublattice $l$ is proportional to
\begin{equation}
\vec{n}^l_{\vec{m}} = (\vec{r}^{\tilde{l}}_{\vec{m}}-\vec{r}^{\tilde{l}}_{\vec{m}^{+}_{l}})\times(\vec{r}^{\tilde{l}}_{\vec{m}}-\vec{r}^{\tilde{l}}_{\vec{m}^{-}_{l}}),
\end{equation}
where cross denotes the vector product. The local dipolar moment is then proportional to
\begin{equation}
\label{eq:moment}
\vec{p}^l_{\vec{m}} = \frac{(\vec{n}^l_{\vec{m}} \cdot (\vec{r}^{l}_{\vec{m}}-\vec{r}^{\tilde{l}}_{\vec{m}}))\vec{n}^l_{\vec{m}}}{|\vec{n}^l_{\vec{m}}|^2},
\end{equation}
where dot stands for the scalar product. Up to the first order in atom displacements $u$ we get $\vec{p}=[0,0,p]$, where $p$ is
\begin{equation}
\label{eq:momentfirst}
p^l_{\vec{m}} = 
u^l_{\vec{m}} - \frac{1}{3} \left( u^{\tilde{l}}_{\vec{m}} + u^{\tilde{l}}_{\vec{m}^{+}_{l}} + u^{\tilde{l}}_{\vec{m}^{-}_{l}}\right).
\end{equation}
The Hamiltonian is then 
\begin{eqnarray}
\label{eq:hamiltonian}
\nonumber H & = & \sum\limits_{l,\vec{m}} \left( \frac{m (a \dot{u}^l_{\vec{m}})^2}{2} + c p^l_{\vec{m}})^2 \right) \\
&+ & b \sum\limits_{l,\vec{m}} \sum\limits_{l',\vec{m}'} 
\frac{p^l_{\vec{m}} p^{l'}_{\vec{m}'}}{|\vec{\delta}|^3} \left. \left(
1 - \frac{3(\vec{e_Z}\cdot\vec{\delta})^2}{|\vec{\delta}|^2}
\right) \right|_{\vec{\delta}=\vec{\rho}^l_{\vec{m}}-\vec{\rho}^{l'}_{\vec{m}'}},
\end{eqnarray}
where $m$ is an atom's mass, $c$ and $b$ are parameters of the model (both have dimensions of energy). Expressing this Hamiltonian in units of $m a^2$, we can introduce two characteristic frequencies $\omega_0$=$c/(m a^2)$ and $\omega_1=\beta \omega_0$ with $\beta=b/c$. The parameter $\omega_0$ defines the overall energy scale (later we normalize it out), while $\beta$ remains the free parameter of the model.

Physically, the model attempts to capture essentials of $\pi$-orbitals polarization during the deformation of each individual graphene sheet. Such deformation produces local shift of the charge from one side of the sheet to another, which can be represented as an additional charge density, superimposed over the original, undeformed, orbital. The first potential energy term in (\ref{eq:hamiltonian}) corresponds to the electrostatic self-energy of this additional density, while the second term models the interaction between these redistributed charges across the whole lattice. This reproduces precisely the extremely short-range (the self-energy, taken simply as an independent parameter) and long-range parts (by keeping the leading-order dipolar terms) of the interaction between deformed orbitals while neglecting the higher-order multipole terms, whose contribution peaks at intermediate distances.

To solve the model one may reexpress the Hamiltonian (\ref{eq:hamiltonian}) in terms of the displacements $u^l_{\vec{m}}$ and differentiate to find the force on an element $\vec{m}$ of each of the four sublattices. Representing the displacements by their Fourier components both in time and space 
\begin{equation}
u^l_{\vec{m}}(t) = \int u^{l}(\vec{k}) e^{2 \pi \imath (\vec{k} \cdot \vec{\rho}^{l}_{\vec{m}})+ \imath \omega t} \ud^3\vec{k},
\end{equation}
where $\vec{k}=[k_X,k_Y,k_Z]$ and the explicit dependence on time $t$ is shown, one  gets the following usual secular equation for the frequency
\begin{equation}
\label{eq:secular}
\omega^2 \left[ 
\begin{array}{c}
u^{1}(\vec{k})\\
u^{2}(\vec{k})\\
u^{3}(\vec{k})\\
u^{4}(\vec{k})\\
\end{array}
\right] =
\frac{\omega_0^2}{9}
\left[ \begin{array}{cccc}
A            & B            & C            & D \\
\overline{B} & A            & E            & C \\
\overline{C} & \overline{E} & A            & B \\
\overline{D} & \overline{C} & \overline{B} & A 
\end{array} \right]
\cdot
\left[ 
\begin{array}{c}
u^{1}(\vec{k})\\
u^{2}(\vec{k})\\
u^{3}(\vec{k})\\
u^{4}(\vec{k})\\
\end{array}
\right],
\end{equation}
where the matrix (called the dynamical matrix and denoted here, including the numerical coefficient 1/9 in front, as $\mathbf{M}$) is obviously self-adjoint. Its elements are
\begin{eqnarray}
\label{eq:A}
A & = & 2 a \left( 2 + \beta \uS_{1} \right)  - 3 \beta \left( \overline{b} \overline{\uS_{2}} + b \uS_{2} \right) \\
\label{eq:B}
B & = & \beta ( \overline{b}^2 \overline{\uS_{2}} + 9 \uS_{2} ) - 6 \overline{b} \left( 2 + \beta \uS_{1} \right) \\
C & = & \beta \left( 3 \overline{b} \uU_{0} - 2 a \uU_{3} + 3 b \uU_{4} \right) \\
D & = & \beta \left( \overline{b} (\overline{b} \uU_{0} - 6 \uU_{3}) + 9 \uU_{4} \right) \\
E& = & \beta \left( 9 \uU_{0} + b ( b \uU_{4} - 6 \uU_{3}) \right),
\end{eqnarray} 
with $\uS_{l} = Z_{\mathbf{b}}(3,\vec{k},\vec{d}_{l})$, $\uU_{l} = \uS_{l} - 3 \alpha^2 Z_{\mathbf{b}}(5,\vec{k},\vec{d}_{l})$, $a=2 \cos \left(\sqrt{3} \pi  k_X \right) \cos (3 \pi 
   k_Y)+\cos \left(2 \sqrt{3} \pi  k_X\right)+6$ and $ b = 1+2 e^{-3 i \pi  k_Y} \cos \left(\sqrt{3} \pi k_X\right)$. This assumes the following definition of the Epstein zeta function
\begin{equation}
\label{eq:zeta}
Z_{\mathbf A}(s, \vec{c}, \vec{d}) = \sum\limits_{\vec{n}\in Z^D}{\!\!}^\prime 
\frac{e^{2 \pi \imath \vec{c}\cdot\mathbf{A}\cdot\vec{n}}}{\left| \mathbf{A}\cdot\vec{n} - \vec{d} \right|^s},
\end{equation}
where prime near the sum means that singular terms are excluded, $\mathbf{A}$ is an arbitrary $D \times D$ matrix, $s$ is (in general) complex scalar and $\vec{c}$, $\vec{d}$ are arbitrary $D$-vectors. The vectors $\vec{d}_l$ are from (\ref{eq:shifts}) with $\vec{d}_0=[0,0,\alpha]$.

Epstein zeta function can be very efficiently evaluated\cite{CrandallZeta98} by a computer program\cite{M09_graphite_math}. The four branches of flexural phonon spectrum of graphite at any point in $\vec{k}$-space are then just the square roots of eigenvalues of the matrix $\mathbf{M}$, defined by (\ref{eq:secular})-(\ref{eq:zeta}). Apart from the parameter $\omega_0$, defining the overall frequency scale, these branches depend on graphite interlayer separation $\alpha = 2.34$, taken from the experiment, and the free parameter of the model $\beta$.

\begin{figure}[t]
  \begin{center}
    \vspace{0.3cm}
    \includegraphics[width=8cm]{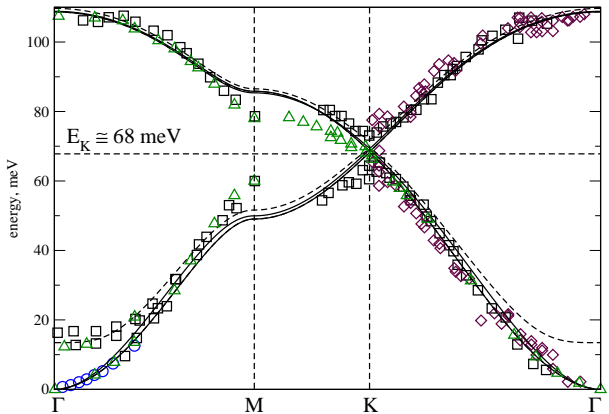}
    \vspace{-0.3cm}
  \end{center}
  \caption{Dispersion of flexural phonons in graphite across the first Brillouin zone, labels correspond to the well known high-symmetry points. Experimental data are shown by circles\cite{NWS72}, squares\cite{YTIMROO05} (as reproduced in Ref.~\onlinecite{MM05}), triangles\cite{MMDRMDBKT07}, diamonds\cite{Siebentritt_PRB2007}. Solid lines are calculated from (\ref{eq:secular})-(\ref{eq:zeta}) with $\beta=0.36$, dashed lines are corrected by (\ref{eq:zoprime}) with $\eta=0.036$.}
  \label{fig:spectr}
\end{figure}

A wealth of experimental data on phonon spectrum of graphite is available in the literature\cite{NWS72,YTIMROO05,MMDRMDBKT07,Siebentritt_PRB2007}. Some of these data are shown in Fig.~\ref{fig:spectr}
along with dispersion curves predicted by the considered model for $\beta=0.36$, giving the best fit to the data. The value of $\omega_0=51 meV$ was obtained by fixing the values of the spectrum at $K$ point. There is a slight disagreement near $M$, which can be attributed either to the oversimplification of the model, neglecting the higher-order multipole terms, or may be even to the experimental errors, which, at least for one set of measurements\cite{NWS72}, increase on approach to $M$. Provided there is a single adjustable parameter, the agreement between the model (solid line) and the experiment is very good.

The spectrum of graphene can be obtained as a limit at $\alpha\rightarrow\infty$. Then $C,D,E\rightarrow0$ (interaction between the layers vanishes) and $\mathbf{M}$ splits into two $2\times2$ sub-matrices. Expressions for $A$ and $B$ remain the same, except that zeta function becomes 2-dimensional as the matrix $\mathbf{b}$ in the expression for $\uS_{l}$ loses its last row and last column. The resulting spectrum is very similar to the one already shown in Fig.~\ref{fig:spectr}, just there is no splitting of acoustic and optical branches. Please note that 
even though the long-range interaction across the layers is eliminated in graphene limit, the interaction inside the layer still remains, giving the spectrum its specific shape.

Let us also note that in the present model displacement of graphite layers as a whole, without flexing them, does not generate any dipolar moment (\ref{eq:moment}) and, thus, leaves the energy invariant. According to the Goldstone theorem, the presence of such continuos symmetry implies the existence of an additional acoustic mode in the spectrum (the other acoustic mode is due to the energy invariance with respect to translation of the whole crystal). In graphite this mode turns optic, acquiring a certain amount of energy at $\Gamma$ point due to macroscopic van der Waals interaction between the carbon layers and depends on their complete flexural phonon spectrum (\ref{eq:secular})-(\ref{eq:zeta}) as well as temperature. This interaction can be simulated by introducing an additional phenomenological harmonic coupling between neighbouring atoms on sublattices $2$ and $3$, resulting in the following addition to the dynamic matrix in (\ref{eq:secular})
\begin{eqnarray}
\label{eq:zoprime}
D'=\eta \left[ \begin{array}{cccc}
0            & 0            & 0            & 0 \\
0            & 2            & -1-e^{4\pi \imath k_Z}          & 0 \\
0 & -1-e^{-4\pi \imath k_Z} & 2    & 0 \\
0 & 0 & 0 & 0 
\end{array} \right],
\end{eqnarray}
where $\eta$ is a parameter. The effect of this addition is mostly localized in the neighbourhood of $\Gamma$ point (see dashed line in Fig.~\ref{fig:spectr}, corresponding to the best-fit value of $\eta=0.036$). In principle, the value of $\eta$ can be calculated on the basis of the present model (by also including the repulsion between the layers due to exchange interaction), but, while its introduction results in a better fit to the data, it is methodologically wrong to try to model the macroscopic interaction inside the microscopic Hamiltonian (\ref{eq:hamiltonian}).

One may try to add $\sigma$-bonds stretching to the model by introducing the following term into the Hamiltonian (\ref{eq:hamiltonian}):
\begin{equation}
 H^\sigma = \frac{\gamma}{2}\sum\limits_{l,\vec{m}}
   \left(
      |\vec{r}^{\tilde{l}}_{\vec{m}^{+}_{l}}-\vec{r}^{l}_{\vec{m}}|^2+
      |\vec{r}^{\tilde{l}}_{\vec{m}^{-}_{l}}-\vec{r}^{l}_{\vec{m}}|^2+
      |\vec{r}^{\tilde{l}}_{\vec{m}_{l}}-\vec{r}^{l}_{\vec{m}}|^2
   \right),
\end{equation}
where $\gamma$ is a free parameter. This models $\sigma$-bonds as harmonic strings with an equilibrium length of $1$, producing, up to the first order in displacements, an additional force $-6\gamma p^l_{\vec{m}}$ at each site (proportionality of this force to the magnitude of the local dipolar moment is just a convenient coincidence). The corresponding dynamical matrix can be obtained by adding $3 \gamma$ to $A$ in Eq. (\ref{eq:A}) and $-\gamma(1 + e^{3 \imath \pi k_Y} \cos \sqrt 3 \pi k_X)$ to $B$ in Eq. (\ref{eq:B}). However, fitting the resulting spectrum to the experimental data, produces (up to the fitting error) $\gamma=0$. It means, there is no contribution of $\sigma$-bonds stretching to flexural phonon spectrum, which may be a consqeuence of their strong anisotropy. There is no need to introduce an additional parameter $\gamma$ to obtain the fit, shown in Fig.~\ref{fig:spectr}.

The interlayer interaction becomes progressively stronger if one presses the graphite, reducing the interlayer distance $\alpha$. In this case splitting of the acoustic and optical branches rapidly grows, until, at a critical value of $\alpha=0.91$, one of the acoustic branches touches the horizontal axis at point $K$. At smaller $\alpha$ the corresponding eigenvalue of $\mathbf{M}$ becomes negative, meaning that the lattice is unstable. This happens at interlayer distance approaching the intra-layer distance between carbon atoms, suggesing that it corresponds to lability boundary of graphite $\rightarrow$ diamond transition.

To conclude, the presented simple model, by explicitly including the long-range dipolar interactions, quantitatively reproduces flexural phonon spectrum of graphite and graphene using a single, universal for all layered carbon allotropes, parameter $\beta$. This is an advantage with respect to the widely used for this task Born-von K\'arm\'an type models, necessitating to include several nearest neighbours (and, consequently, many parameters) into consideration to obtain comparable agreement to the experiment. Because the interactions in the presented model are purely electrostatic, their dependence on inter-atomic and inter-layer distance is explicit. The model is not specific to graphene or graphite. The expression for stress-induced dipolar moment and a similar Hamiltonian may be useful while considering other layered carbon allotropes, such as single- and multi- walled nano-tubes, fullerenes etc. 

The effect of higher order multipoles or other quickly decaying short-range interactions would correspond to a certain simple addition to the dynamical matrix, similar to discussed above. Since the most important (and more difficult to evaluate) long-range contribution is already taken into account (and produces a very complete-looking spectrum), one can expect the other contributions to be small and local. Thus, it can be hoped that the model can be a solid basis for further quantitative improvement.



\end{document}